\documentclass[journal]{IEEEtran}
\usepackage{graphicx} 
\usepackage{xcolor}
\usepackage{amsmath}
\usepackage{amssymb}
\usepackage{comment}

\DeclareMathOperator\sign{sign}
\DeclareMathOperator*{\argmin}{arg\,min}

\begin{document}
\title{Performance Optimization of Short Reach Optical Interconnects based on Direct Detection}

\author{Luca~Potì,~\IEEEmembership{Senior Member,~IEEE,} Stella Civelli,~\IEEEmembership{Member,~IEEE,} Marco Secondini,~\IEEEmembership{Senior Member,~IEEE,} Li Zhang, Dario Cellini, Asfand Nizamani, Ramin Solaimani,~\IEEEmembership{Member,~IEEE,} Lorenzo De Marinis, Mareli Rodigheri, Pantea Nadimi Goki,~\IEEEmembership{Member,~IEEE,} Muhammad A. Naz,  Giampiero Contestabile~\IEEEmembership{Senior Member,~OPTICA,} Enrico Forestieri,~\IEEEmembership{Life Member,~IEEE,} and Fabio Cavaliere.
\thanks{L. Potì is with Consorzio Nazionale Interuniversitario per le Telecomunicazioni (CNIT)- Photonic Networks and Technologies Laboratory (PNTLab), Pisa and Universitas Mercatorum, Rome - Italy e-mail: luca.poti@cnit.it; A. Nizamani and M. A. Naz were with CNIT-PNTlab, Pisa - Italy; L. Zhang and R. Solaimani are with CNIT-PNTlab, Pisa - Italy; D. Cellini, L. De Marinis, P. Nadimi Goki, M. Secondini, G. Contestabile, and E. Forestieri are with Scuola Superiore Sant'Anna, Pisa - Italy; M. Rodigheri was with Universidade Estadual de Campinas, Campinas, SP - Brazil; S. Civelli is with CNR - Italy; F. Cavaliere is with Ericsson Research, Pisa - Italy.}}
\IEEEspecialpapernotice{(Invited Paper)}
\maketitle

\begin{abstract}
Short-reach optical interconnects are evolving toward data rates beyond 400 Gb/s per lane, driven by the bandwidth and energy-efficiency requirements of AI-enabled datacenter networks. At these operating speeds, channel impairments, device nonlinearities, and hardware constraints limit the effectiveness of conventional transceiver design and digital signal processing (DSP). This paper presents a unified framework for the optimization of direct-detection optical interconnects, encompassing digital surrogate modeling, receiver-side DSP optimization, and end-to-end (E2E) transceiver learning. The proposed formulation provides a common perspective for model-based and machine-learning-based approaches, including linear and nonlinear equalization, lookup tables, decision trees, neural-network receivers, and E2E optimization. Their performance is discussed together with computational complexity and hardware implementation aspects, highlighting the associated trade-offs. We show, through simulations and experimental validations, that for a 40~Gb/s 10~km link, decision trees can outperform by 0.5--1~dB conventional linear equalization, with negligible hardware requirements. Moreover, we show that an E2E technique based on transformers can provide a gain up to 6~dB, highlighting the potential of joint transceiver optimization to improve the performance of next-generation short-reach optical links.\end{abstract}

\IEEEpeerreviewmaketitle

\section{Introduction}

\IEEEPARstart{T}{he} rapid evolution of high-speed interconnects is driven by the exponential scaling of bandwidth demands in AI-driven datacenter and communication systems. Industry standardization bodies, notably IEEE 802.3 and the optical internetworking forum (OIF), have progressively pushed electrical and optical interconnect technologies toward 400~Gb/s per lane and beyond. This evolution is closely tied to the requirements of large-scale AI infrastructures, where both scale-up and scale-out networks demand unprecedented bandwidth density, energy efficiency, and signal integrity~\cite{Ghiasi2026}.

Within IEEE 802.3, ongoing efforts focus on defining next-generation Ethernet physical layers capable of supporting 400~Gb/s per lane signaling, motivated by the need to sustain increasingly dense accelerator interconnects and support media access control (MAC) rates beyond 1.6~Tb/s. In parallel, the OIF is advancing electrical interface standards (e.g., CEI-224G and CEI-448G) and investigating higher-capacity solutions, including 6.4~Tb/s and 12.8~Tb/s co-packaged optics (CPO) modules, highlighting a strong convergence between electrical and optical interface standardization. These developments reflect a broader trend toward tighter integration between compute and optics, including emerging paradigms such as linear pluggable optics (LPO), near-packaged optics (NPO), and CPO~\cite{Ghiasi2026}.

However, as interconnect speeds approach and exceed 400~Gb/s per lane, traditional signal processing techniques based on linear equalization and conventional digital signal processing (DSP) face fundamental limitations. The increased impact of channel loss, crosstalk, and bandwidth constraints necessitates aggressive equalization strategies, which in turn amplify noise and significantly increase power consumption. Moreover, standard modeling approaches such as channel operating margin (COM), originally developed for linear copper links, rely on assumptions of linearity and time invariance that become increasingly inaccurate in the presence of nonlinearities introduced by optical components and advanced modulation formats. 
These challenges are further exacerbated in emerging interconnect architectures such as LPO and CPO, where DSP functionality is partially or fully redistributed or minimized to improve energy efficiency. In such architectures, equalizers implemented in the host serializer/deserializer (SerDes) must compensate not only for electrical impairments (e.g., printed circuit board, PCB, loss and crosstalk) but also for optical impairments (e.g., limited bandwidth and nonlinear behavior of transmitters and receivers). This creates a complex, highly coupled optimization problem across the entire link, where traditional block-by-block design approaches become suboptimal or inadequate. 

Beyond conventional block-by-block transceiver design, several optimization strategies have been proposed to improve the performance of short-reach optical links. Depending on the available system knowledge and the targeted subsystem, optimization may involve the operating conditions of individual optoelectronic devices, transmitter (TX)-side digital pre-distortion, receiver (RX)-side equalization and detection, or the joint optimization of TX and RX processing. These approaches range from analytical model-based techniques, such as minimum mean-square error equalization, Volterra filtering and maximum-likelihood detection, to data-driven methods based on neural networks (NNs) and differentiable optimization, each offering different trade-offs between performance, complexity and implementation cost~\cite{Smith1965,Jang1998,Stojanovic2017,Che2024,Karanov2018}.
Rather than focusing on a single optimization strategy, this paper presents a unified framework encompassing digital surrogate modeling, RX-side optimization, and end-to-end (E2E) transceiver optimization for short-reach optical interconnects. Within this framework, conventional model-based algorithms and recent machine-learning approaches are described using a common formulation, enabling a systematic comparison of their performance, computational complexity, and hardware implementation requirements. Beyond providing a unified methodological perspective, this paper offers a comprehensive assessment of multiple optimization solutions from an implementation-oriented standpoint, explicitly analyzing their computational and hardware complexity and supporting the comparison through experimental validation.
\begin{figure*}
\centering
\begin{centering}
\includegraphics[width=1.5\columnwidth]{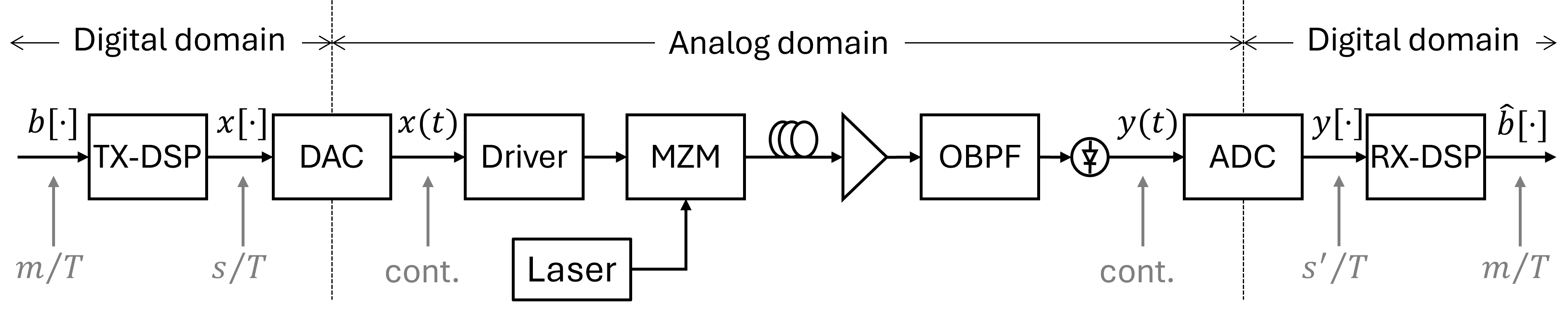}
\par\end{centering}
\caption{System overview. Discrete-time sequences are denoted with empty brackets, e.g., $x[\cdot]$, and the corresponding rate is reported below in gray. }\label{fig:system_overview}

\end{figure*}
The remainder of this paper is organized as follows. Section~II introduces the considered optical communication system and establishes the notation adopted throughout the paper. Section~III presents a unified framework for transceiver optimization based on digital surrogate modeling, RX-side optimization, and E2E learning. Section~IV discusses differentiable digital surrogate models for the main analog components of the optical link. Section~V describes representative transceiver optimization techniques, including RX-side and E2E approaches and discusses the computational complexity and hardware implementation aspects of the considered algorithms. Section~VI presents both simulation and experimental results. Finally, Section~VII concludes the paper.

\section{System Overview}\label{sec:System-overview}



We consider the generic optical communication system illustrated in
Fig.~\ref{fig:system_overview}. The TX
comprises a DSP (TX-DSP) unit followed by a
digital-to-analog converter (DAC), an electrical driver, a laser source,
and an optical Mach--Zehnder modulator (MZM). The MZM modulates the optical field so that the transmitted optical waveform is, in general, characterized by both amplitude and phase. Although the RX employs direct detection, and short-reach systems are therefore commonly referred to as intensity-modulation/direct-detection (IM/DD) systems, the transmitted optical field may still carry useful phase information that can be indirectly recovered through suitable RX processing. For this reason, throughout this paper we adopt the more general term amplitude-modulation/direct-detection (AM/DD) to emphasize that the transmitted optical field is not restricted to conventional intensity modulation. BPAM-4, discussed in Section~\ref{subsec:BPAM}, provides a representative example of this principle.

The optical signal
propagates through the fiber link, is filtered through an optical band-pass filter (OBPF) to remove out-of-band noise and detected by a photodetector at the RX. The resulting electrical signal is digitized
by an analog-to-digital converter (ADC) and processed by the RX
DSP (RX-DSP), which estimates the transmitted information bits.

We distinguish between digital and analog subsystems. Digital blocks,
namely the TX-DSP and RX-DSP, implement programmable algorithms that
can be modified and optimized almost arbitrarily. In contrast, analog
blocks---including the DAC, ADC, electrical and optical devices,
and the propagation channel---are characterized by fixed physical
transfer functions that may be only partially known and generally
impose non-idealities and implementation constraints on the overall
system. The DAC and ADC constitute the interface between the digital
and analog domains.

The main objective of the transceiver design is to determine the TX-DSP
and RX-DSP functions that maximize the communication performance over
the given analog subsystem. From this perspective, the modulation
format, pulse shaping, equalization, and detection strategy are not
predefined but become design variables that can be jointly optimized.

The sequence of information bits $b[\cdot]$ is transmitted at a rate
of $m$ bits per symbol period $T$, with corresponding bit rate $R_{b}=m/T$,
so that $b[n]$ is the bit transmitted at time $t=nT/m$. For convenience,
the bits transmitted during the $k$-th symbol interval are collected
into the binary vector 
\begin{equation}
\mathbf{b}[k]=\left(b[mk],\ldots,b[mk+m-1]\right)^{T}\in\{0,1\}^{m}
\end{equation}

The TX-DSP maps the input bit sequence into the discrete-time waveform
$x[\cdot]$ represented by $s$ samples per symbol, so that $x[n]$
denotes the sample generated at time $t=nT/s$. The vector 
\begin{equation}
\mathbf{x}[k]=\left(x[sk],\ldots,x[sk+s-1]\right)^{T}\in\mathbb{R}^{s}
\end{equation}
collects the $s$ samples generated during the $k$-th symbol interval.

The DAC converts the digital sequence into the continuous-time electrical
waveform $x(t)$. Under an ideal DAC model, infinite resolution and
ideal interpolation with a sinc function (rectangular frequency response)
are assumed. Practical DACs, however, are affected by finite amplitude
resolution, non-ideal interpolation filtering, and other hardware
impairments. These impairments will be considered in the following
sections.

After propagation through the analog electrical-optical-electrical
subsystem, the received continuous-time waveform $y(t)$ is sampled
by the ADC at sampling rate $s'/T$, obtaining the sample sequence
$y[\cdot],$ with $y[n]=y(nT/s')$. In analogy with the TX-side representation,
the samples received during the $k$-th symbol interval are collected
in the vector
\begin{equation}
\mathbf{y}[k]=\left(y[s'k],\ldots,y[s'k+s'-1]\right)^{T}\in\mathbb{R}^{s'}
\label{eq:y_vec}
\end{equation}
Similarly to the DAC, an ideal ADC is modeled as an ideal anti-aliasing
filter with a rectangular frequency response, followed by uniform
sampling and quantization with infinite precision. In practice, ADCs
are affected by finite quantization resolution, non-ideal analog front-end
filtering, and other hardware impairments.

Finally, the RX-DSP processes the received sample sequence to provide
an estimate of the transmitted bit sequence $\hat{b}[\cdot]$. The
bits estimated during the $k$-th symbol interval are collected in
the vector
\begin{equation}
\hat{\mathbf{b}}[k]=\left(\hat{b}[mk],\ldots,\hat{b}[mk+m-1]\right)^{T}\in\{0,1\}^{m}
\end{equation}

\section{Modeling and Optimization}\label{sec:Modeling-and-optimization}

Several approaches can be adopted for the design of the transceiver,
depending on the complexity of the communication channel and the available
knowledge of the underlying physical system.

The conventional approach consists of selecting a priori a modulation
format, pulse-shaping filter, sampling strategy, and detection algorithm.
The corresponding TX-DSP and RX-DSP are then designed according to
the selected modulation format and the system constraints. This methodology
is particularly effective for relatively simple channels, such as
bandwidth-limited additive white Gaussian noise (AWGN) channels, for
which near-optimal modulation and detection techniques are well established.
Examples include conventional pulse-amplitude modulation (PAM) formats,
as well as the recently proposed bipolar PAM (BPAM), which exploits
the observation that, even in direct-detection systems, encoding information
onto bipolar electrical waveforms may be feasible and, under suitable
conditions, advantageous.

In many practical scenarios, however, the overall electro-optical-electrical
link, such as that illustrated in Fig.~\ref{fig:system_overview},
deviates significantly from the ideal AWGN model because of device
imperfections, nonlinearities, memory effects, and partially unknown
or time-varying characteristics. A first strategy is therefore to
preserve the selected modulation format while optimizing the RX-DSP
to improve the reliability of the detected bits. In this case, the
RX is described by a generic parameterized processing structure,
whose parameters are learned by minimizing a suitable loss function
on representative training data. Depending on the application, the
adopted model may range from conventional adaptive feed-forward equalizers
(FFE) trained according to the minimum mean-square error criterion,
to more powerful nonlinear structures such as Volterra equalizers,
memory polynomials, or deep NNs. Alternatively,
low-complexity parameterized models, including LUTs and decision trees,
may be employed when implementation cost is a primary concern.

The most general approach consists of jointly optimizing both the
TX-DSP and RX-DSP for the target communication channel. The underlying
rationale is that, for sufficiently complex channels, the optimum
transmitted waveform may differ substantially from conventional modulation
formats originally developed for AWGN channels. The BPAM example discussed
above provides a simple illustration of this principle, showing that, somewhat surprisingly, encoding information in the signal sign can provide a significant performance gain even in the presence of a squaring operation. This design
paradigm, commonly referred to as E2E optimization, has recently
attracted considerable attention thanks to advances in machine learning
and differentiable optimization. Although more demanding than RX-only
optimization, it offers substantially greater design flexibility by
jointly optimizing the TX and RX as a single communication
system. The remainder of this section focuses on this general framework,
while conventional transceiver design and RX-only optimization
will be regarded as particular, constrained instances of the same
methodology.

\begin{figure}
\centering
\begin{centering}
\includegraphics[width=1\columnwidth]{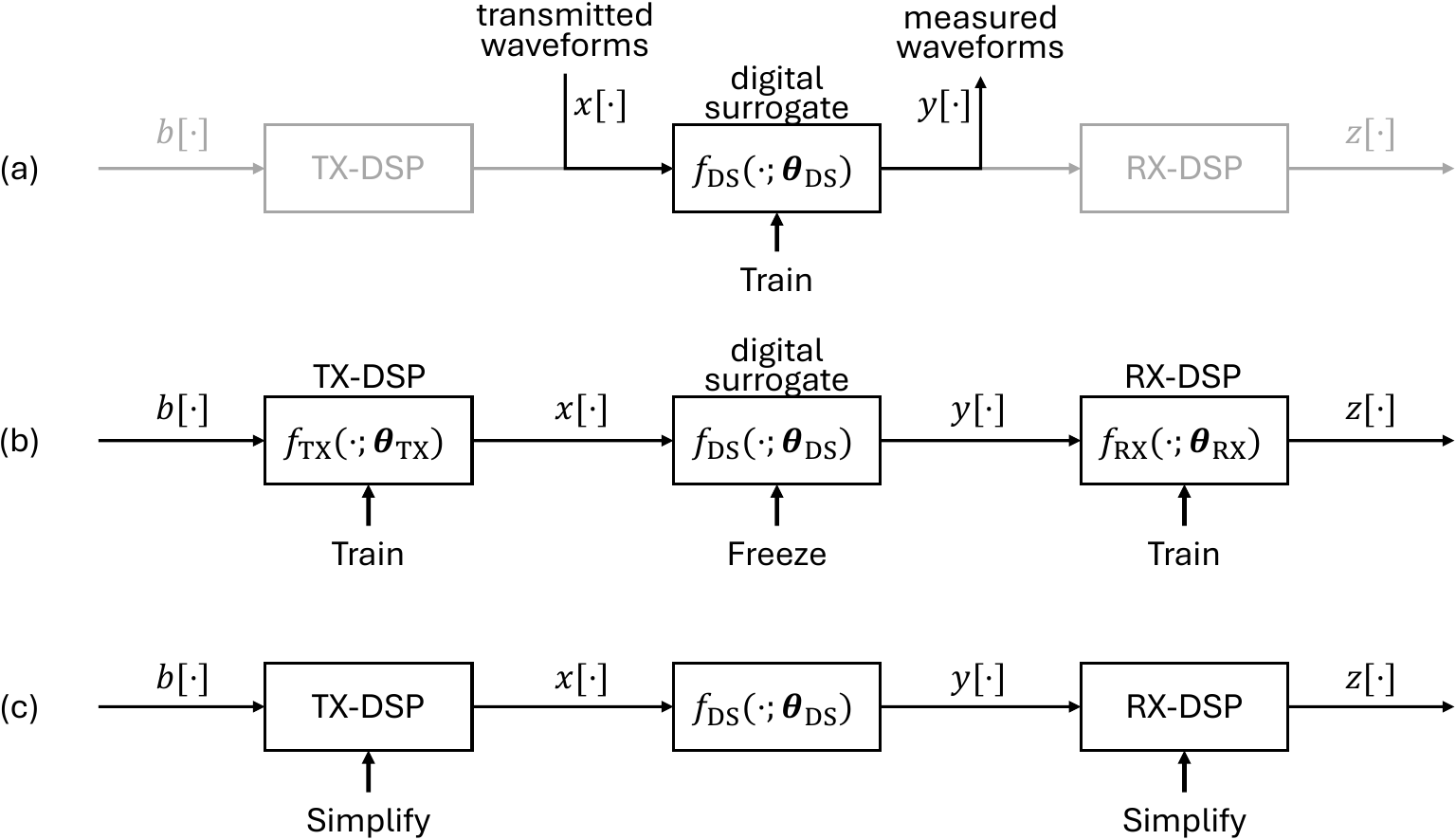}
\par\end{centering}
\caption{Three-step procedure for E2E modeling and optimization: (a)
digital surrogate modeling; (b) E2E transceiver optimization;
and (c) DSP simplification. }\label{fig:system_modelling_optimization}
\end{figure}

The proposed framework consists of the three-step procedure illustrated
in Fig.~\ref{fig:system_modelling_optimization}: (a) \emph{digital
surrogate modeling}, (b) \emph{E2E DSP optimization}, and (c)
\emph{DSP simplification}. The first step identifies a differentiable
model of the analog subsystem from input--output data. The resulting
digital surrogate (DS) is then used in the second step to jointly
optimize the TX- and RX- DSP algorithms through gradient-based
E2E learning. 
The final step aims at reducing the implementation complexity of the optimized DSP by exploiting the structure of the learned mappings, replacing complex neural-network models with functionally similar but more hardware-friendly implementations, such as lookup tables (LUTs) when the input space is sufficiently small or decision trees when the learned mapping can be accurately represented by piecewise decision boundaries.
The first two steps are described in detail in the following two subsections, while the simplification techniques are discussed in the subsequent sections presenting the corresponding practical DSP implementations.

\subsection{Digital Surrogate Modeling}

The objective of the DS is to reproduce the input--output behavior
of the complete analog subsystem, including the DAC, electrical and
optical devices, transmission channel, photodetector, and ADC. The
DS is trained from one or more transmitted and received waveforms,
acquired either experimentally from the physical system or generated
by high-fidelity numerical simulations, as shown in Fig.~\ref{fig:system_modelling_optimization}(a).

\begin{figure}
\begin{centering}
\includegraphics[width=1\columnwidth]{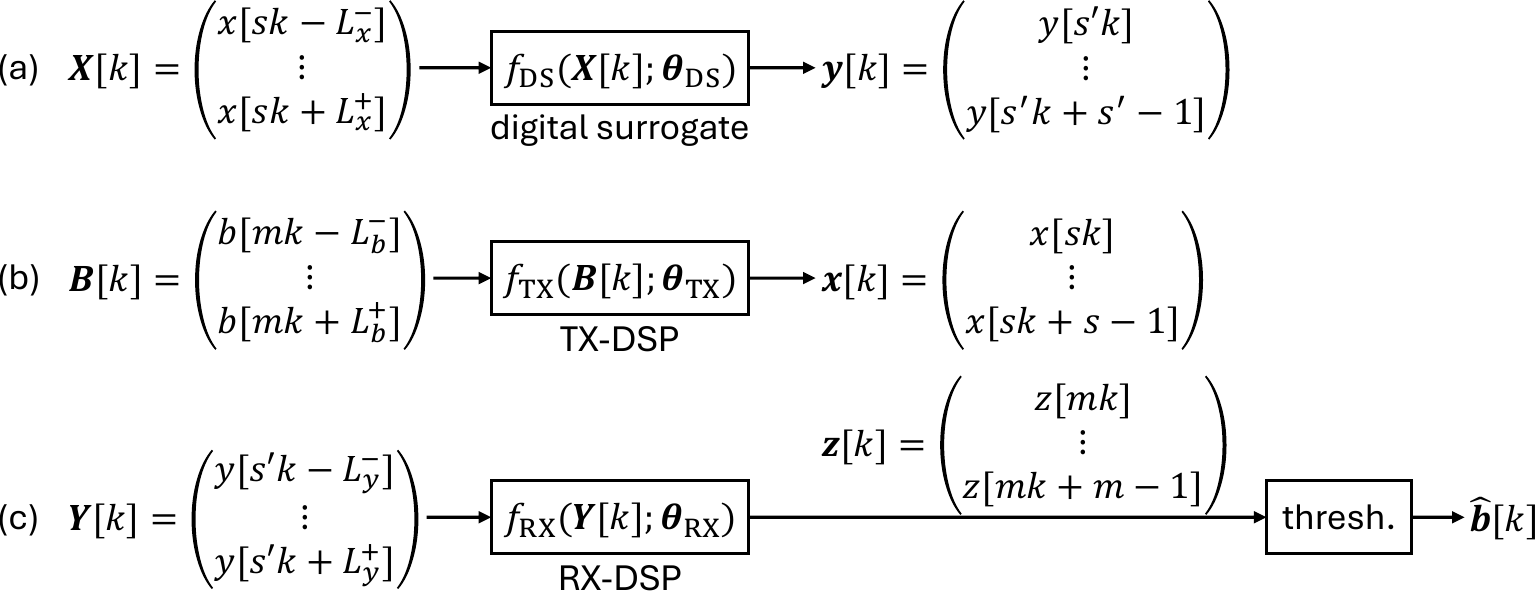}
\par\end{centering}
\caption{Input--output representation of the differentiable mappings employed
in the proposed framework: (a) digital surrogate (DS), (b) TX-DSP,
and (c) RX-DSP. Vectors $\boldsymbol{\theta}_{\mathrm{DS}}$, $\boldsymbol{\theta}_{\mathrm{TX}}$,
$\boldsymbol{\theta}_{\mathrm{RX}}$ denote the corresponding trainable
parameters.}\label{fig:input_output_functions}
\end{figure}

Assuming that the analog subsystem has finite memory, spanning $L_{x}^{-}$
precursor and $L_{x}^{+}$ postcursor sampling intervals, we introduce
a sliding context window of length $L_{x}=L_{x}^{-}+L_{x}^{+}+1$
samples, shifted by $s$ samples (i.e., one symbol interval) at each
step. The input samples contained in the $k$-th window are collected
in the vector\footnote{Three notations are used to distinguish different temporal organizations
of the same quantities: lowercase symbols (e.g., $b[n]$, $x[n]$,
$y[n]$) refer to sequences at the bit or sampling rate, bold lowercase
symbols (e.g., $\mathbf{b}[k]$, $\mathbf{x}[k]$, $\mathbf{y}[k]$)
to symbol-rate vectors grouping the samples (or bits) within the $k$-th
symbol interval, and bold uppercase symbols (e.g., $\mathbf{B}[k]$,
$\mathbf{X}[k]$, $\mathbf{Y}[k]$)) to symbol-rate context-window
vectors collecting the samples (or bits) within the $k$-th sliding
window.}
\begin{equation}
\mathbf{X}[k]=\left(x[sk-L_{x}^{-}],\ldots,x[sk+L_{x}^{+}]\right)^{T}\in\mathbb{R}^{L_{x}}
\end{equation}
associated with the $k$-th transmitted symbol. The DS  models the
analog subsystem according to
\begin{equation}
\mathbf{y}[k]=f_{\mathrm{DS}}(\mathbf{X}[k];\boldsymbol{\theta}_{\mathrm{DS}})\label{eq:f_DS}
\end{equation}
where the vector $\boldsymbol{\theta}_{\mathrm{DS}}$ collects all
the trainable parameters of the model. The corresponding input--output
mapping is illustrated in Fig.~\ref{fig:input_output_functions}(a).

The function $f_{\mathrm{DS}}$ must be differentiable with respect
to both its parameters and its input in order to enable gradient-based
optimization during both surrogate training and subsequent E2E
optimization. In practice, differentiable deep NN models provide a
flexible solution capable of approximating nonlinear systems with
memory. Since the DS is employed only during the offline design phase,
its computational complexity is generally of secondary importance,
whereas modeling accuracy is crucial. Nevertheless, excessively complex
models may require large training datasets and lead to slow convergence
or overfitting. A promising alternative is offered by physics-inspired
NN, in which the network architecture explicitly reflects the underlying
physical system. Instead of relying on generic trainable layers, individual
network blocks are designed to reproduce the functionality of specific
analog components. For example, linear time-invariant filters can
be implemented as linear layers whose trainable coefficients represent
the filter impulse response, while a photodetector can be modeled
as a quadratic activation followed by a linear filtering stage. In
this way, only the unknown physical parameters are learned from the
data, leading to more interpretable models and often improving both
training efficiency and generalization. Further examples will be discussed
in the following sections.

The surrogate parameters $\boldsymbol{\theta}_{\mathrm{DS}}$ are
estimated by supervised learning using a suitable loss function, such
as the mean-square error between the predicted and measured output
waveforms, and optimized with standard gradient-based algorithms.

The analog subsystem may be modeled either by separate surrogates
describing individual components or by a single E2E model.
In practice, however, obtaining separate input-output training datasets
for each component is often impractical or even impossible. A convenient
compromise is to adopt a modular physics-inspired architecture, in
which each component model is individually designed and validated
in simulation, while the complete DS is finally trained E2E
using experimental waveform data.

When non-differentiable operations such as quantization are included,
suitable differentiable approximations can be adopted during training;
these will be discussed in the following sections.

\subsection{End-to-end (E2E) Optimization}\label{subsec:End-to-end-optimization}

Once the DS has been trained, its parameters $\boldsymbol{\theta}_{\mathrm{DS}}$
are frozen and the DS is incorporated into the E2E communication
chain, as shown in Fig.~\ref{fig:system_modelling_optimization}(b).
The TX-DSP and RX-DSP blocks are then jointly trained in an autoencoder-like
framework, with the objective of recovering at the RX output the same
information bits applied to the TX input after propagation through
the communication system modeled by the frozen DS.

To limit the computational complexity, the TX-DSP is assumed to have
finite memory spanning $L_{b}^{-}$ precursor and $L_{b}^{+}$ postcursor
bits. Let
\begin{equation}
\mathbf{B}[k]=\left(b[mk-L_{b}^{-}],\ldots,b[mk+L_{b}^{+}]\right)^{T}\in\{0,1\}^{L_{b}}
\end{equation}
denote the bits contained in a context window of length $L_{b}=L_{b}^{-}+L_{b}^{+}+1$
during the $k$-th symbol interval. The corresponding $s$ samples
of the digital waveform generated by the TX-DSP are given by 
\begin{equation}
\mathbf{x}[k]=f_{\mathrm{TX}}(\mathbf{B}[k];\boldsymbol{\theta}_{\mathrm{TX}})\label{eq:f_TX}
\end{equation}
where $\boldsymbol{\theta}_{\mathrm{TX}}$ collects the trainable
TX parameters. Therefore, the mapping is applied through a sliding
context window of $L_{b}$ bits that advances $m$ bits (i.e., one
symbol interval) at a time and is illustrated in Fig.~\ref{fig:input_output_functions}(b).

Similarly, the RX-DSP operates on a finite context window spanning
$L_{y}^{-}$ precursor and $L_{y}^{+}$ postcursor samples. Defining
\begin{equation}
\mathbf{Y}[k]=\left(\mathbf{y}[s'k-L_{y}^{-}],\ldots,\mathbf{y}[s'k+L_{y}^{+}]\right)^{T}\in\mathbb{R}^{L_{y}}\label{eq:cont_win}
\end{equation}
with $L_{y}=L_{y}^{-}+L_{y}^{+}+1$, the RX estimates the a
posteriori probabilities of the transmitted bits, 
\begin{equation}
\mathbf{z}[k]=\left(z[mk],\ldots z[mk+m-1]\right)^{T}\in[0,1]^{m}
\end{equation}
where
\begin{equation}
z[mk+i]=P(b[mk+i]=1|\mathbf{Y}[k]),\quad i=0,\ldots m-1
\end{equation}
through the mapping
\begin{equation}
\mathbf{z}[k]=f_{\mathrm{RX}}(\mathbf{Y}[k];\boldsymbol{\theta}_{\mathrm{RX}})\label{eq:f_RX}
\end{equation}
where $\boldsymbol{\theta}_{\mathrm{RX}}$ collects the RX parameters.
Soft outputs are particularly suitable for training and for subsequent
soft-input forward error correction (FEC) decoding. Hard decisions
can be readily obtained by thresholding the posterior probabilities
according to
\begin{equation}
\hat{b}[n]=\begin{cases}
1, & z[n]\ge1/2\\
0, & z[n]<1/2
\end{cases}
\end{equation}
The corresponding input--output mapping is illustrated in Fig.~\ref{fig:input_output_functions}(c).

\begin{figure}
\begin{centering}
\includegraphics[width=1\columnwidth]{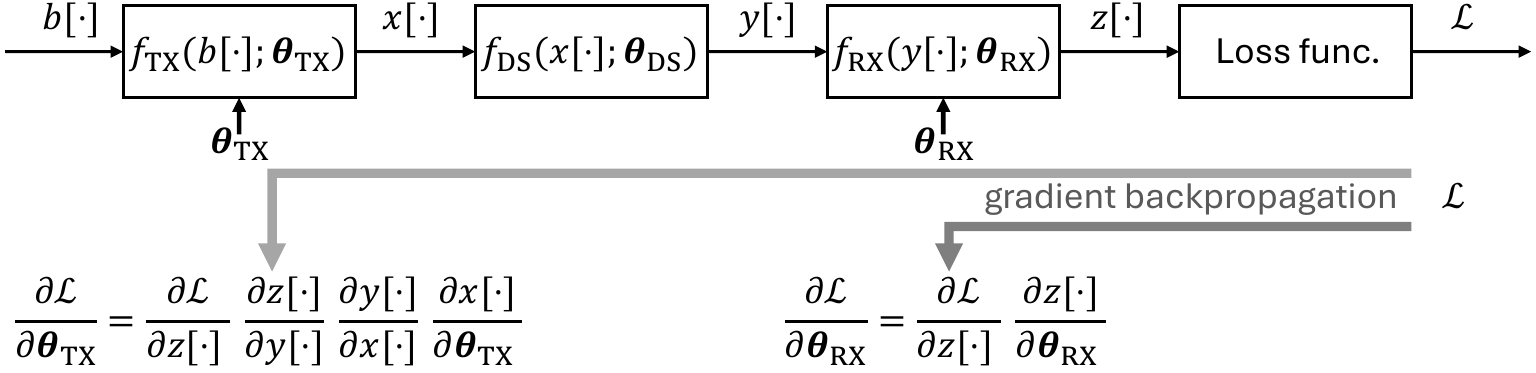}
\par\end{centering}
\caption{Gradient backpropagation for E2E optimization.}
\label{fig:gradient_backpropagation}
\end{figure}

The TX and RX parameters, $\boldsymbol{\theta}_{\mathrm{TX}}$ and
$\boldsymbol{\theta}_{\mathrm{RX}}$, are jointly optimized using
an autoencoder-based self-supervised
learning framework. The binary cross-entropy between the transmitted
bits and their estimated posterior probabilities is adopted as a loss
function {$\mathcal{L}$}. The corresponding optimization problem is solved using standard
gradient-based algorithms, requiring the gradient of {$\mathcal{L}$}
with respect to all trainable parameters to be computed by backpropagation
through the entire communication chain. Fig.~\ref{fig:gradient_backpropagation}
illustrates the resulting gradient backpropagation process. 
The notation $\partial y[\cdot]/\partial x[\cdot]$
denotes the derivative (Jacobian) of the mapping from the whole sequence
$x[\cdot]$ to the whole sequence $y[\cdot]$. Since each processing
block operates on a finite context window, the corresponding Jacobians
are banded, ensuring that the computational complexity of gradient
evaluation remains bounded.

 The function $f_{\mathrm{RX}}$ must be differentiable with respect
to both its inputs and its parameters, whereas $f_{\mathrm{DS}}$
must be differentiable with respect to its inputs (and, during DS
training, also with respect to its parameters). Finally, $f_{\mathrm{TX}}$
must be differentiable only with respect to its trainable parameters.
Suitable differentiable implementations of the various processing
blocks will be described in the following sections. The RX-only
optimization problem is considerably simpler, since the gradient does
not need to backpropagate through the analog subsystem, and will be
specifically discussed in Section~\ref{subsec:Receiver-DSP}.
The framework described above is general and assumes that the differentiable representation of the analog subsystem may be identified from input--output data through the training of $(f_{\mathrm{DS}})$. In the remainder of this paper, however, we consider the specific case in which such a trained surrogate is not yet available and instead employ differentiable, physics-based component models whose parameters are directly determined from analytical descriptions or independent component characterization.

\section{Differentiable Models of the Analog Subsystem}\label{sec:analog-modeling}

Differentiable models (DMs) are here derived from physics-based descriptions of the individual analog components. The resulting model preserves the same input--output structure and differentiability requirements as $(f_{\mathrm{DS}})$, but its parameters are obtained from analytical models or independent component characterization and are not jointly trained from E2E waveform data. The analog subsystem is therefore represented as a cascade of per-component DMs each mirroring the corresponding physical block.

Following the physics-inspired approach  introduced there, the analog subsystem is decomposed
into a chain of per-component DMs, each mirroring one physical block, so that only the unknown parameters of that block are learned. 

In the following, each component is described as a mapping from its input
sample sequence $u[n]$ to its output sample sequence $v[n]$, the output of one component being the input of the next, so that $u[n]=x[n]$ at the input of the first component and $v[n]=y[n]$ at the output of the last.

\subsection{Digital-to-analog (DAC) converter}\label{subsec:dac}
The DAC is modeled as a two-step process. First, the input samples $x[\cdot]$ are mapped to quantized samples with $N_{\mathrm{DAC}}$ bits, then these are converted to the continuous time electrical waveform $x(t)$ using an ideal analog filter. 

We adopt the mid-tread characteristic ~\cite{GershoGray1992} for both the DAC and the ADC.
Being piecewise constant, the quantizer has zero derivative almost everywhere
and an undefined derivative at its transitions. Inserted directly into the
model, it would null the gradient and sever the TX from the
computational graph during backpropagation. It is therefore replaced, during
training, by the straight-through
estimator~\cite{Rokh2023,Jovanovic2023}. In the forward pass the true
quantizer is applied. The model therefore operates on the actual finite-resolution
signal and the effect of the converter resolution is retained. In the backward
pass its derivative is taken to be that of the identity. The gradient then
passes through the quantizer unchanged and reaches the preceding stages. The
identity is the standard choice for this estimator and is what renders the
finite-resolution operation trainable.
The converter also introduces memory, through the analog output filter, together
with a mild nonlinearity of the conversion. Where these effects are significant,
they could be represented by a memory polynomial~\cite{Ding2004}.


Other converter impairments, including timing jitter, integral and differential
    nonlinearity, and thermal noise, are neglected with respect to the ASE introduced by the optical amplifier at the receiver.

\subsection{Electro-optic modulator}\label{subsec:mzm}
The electro-optic modulator is modeled as a single-drive MZM biased at the null
point. Denoting by $V[n]$ the drive voltage and by $V_{\pi}$ the half-wave
voltage, the normalized drive is $u[n]=V[n]/V_{\pi}$, and the mapping from drive
samples to the normalized optical field samples is~\cite{Secondini2020}
\begin{equation}
v[n]=\left[\sin\left(\frac{\pi}{2}u[n]\right)
-j\,\frac{1}{\sqrt{\varepsilon_{r}}}\cos\left(\frac{\pi}{2}u[n]\right)\right]e^{j\frac{\pi}{2}u[n]}
\label{eq:mzm}
\end{equation}
where $\varepsilon_{r}$ is the static extinction ratio (ER). The common phase factor $e^{j\pi u[n]/2}$ is the drive-dependent chirp of the single-drive
configuration. Apart from that factor, the characteristic is odd in the drive up
to the quadrature term, so that drives of opposite sign produce fields of
opposite sign and the antipodal structure of the transmitted signal is preserved
through the modulator. The mapping is analytic in $u[n]$ being $V_{\pi}$ and $\varepsilon_{r}$ fixed device properties.

\subsection{Optical noise}\label{subsec:noise}

Amplified spontaneous emission (ASE) from the optical amplification stage following the fiber is modeled as AWGN on the signal, so that the output of this stage is
\begin{equation}
v[n]=u[n]+w[n],\qquad w[n]\sim\mathcal{N}\!\left(0,\sigma^{2}\right)
\label{eq:noise}
\end{equation}
where the samples $w[n]$ are independent, zero-mean, and Gaussian, with a flat
spectrum across the modeled bandwidth. The variance $\sigma^{2}$ is set so that the resulting signal-to-noise ratio, measured over a reference optical bandwidth as on the
optical spectrum analyzer, matches the target optical signal-to-noise ratio
(OSNR). Within the DM the noise term is a stochastic node
carrying no trainable parameters. Since $w[n]$ is drawn independently of the
signal, the derivative of $v[n]$ with respect to $u[n]$ is unity, so the gradient traverses the node unchanged and, unlike quantization, no surrogate is
required in the backward pass. A new realization is drawn at each forward pass,
so that the optimization is exposed to the noise statistics rather than to a
single realization, while the variance enters as a fixed hyperparameter
set by the target OSNR.

\subsection{Optical filter}\label{subsec:optfilt}

The optical filtering is performed by the OBPF placed before the photodetector. The device is a liquid crystal on silicon (LCoS)-based wavelength-selective switch and is modeled as a bandpass filter
acting on the complex field. Its transfer function is that of a flat-top passband with Gaussian edges as reported in ~\cite{Pulikkaseril2011} eq. (5). The expression is the convolution of a rectangular passband of width $B$ with a Gaussian of 3-dB bandwidth $B_{\mathrm{otf}}$, so that a single function reproduces both the flat top and the finite edge slope of the device.

\subsection{Photodetector}\label{subsec:pd}
The photodetector is modeled as a quadratic activation followed by a linear filtering stage. The square-law activation
\begin{equation}
v[n]=\bigl|u[n]\bigr|^{2}
\label{eq:pd}
\end{equation}
converts the complex field into the photocurrent $v[n]$ and is the essential nonlinearity of the link. The mapping is analytic, differentiable, and introduces no learned parameter. Shot and thermal noise are neglected.

\subsection{Analog-to-digital (ADC) converter}\label{subsec:adc}
The ADC samples $u[n]$ at rate $s'/T$ and quantizes each sample to
$N_{\mathrm{ADC}}$ bits, yielding the received vector $\mathbf{y}[k]$
of~\eqref{eq:y_vec}. Sampling is a differentiable decimation. The quantization
follows the same mid-tread characteristic adopted for the DAC, now at resolution
$N_{\mathrm{ADC}}$, and is likewise handled by the straight-through estimator.

\subsection{Post-detection electrical filter}\label{subsec:postfilt}
The post-detection electrical filter is modeled as a low-pass Gaussian filter with $B_p$ bandwidth. It rejects out-of-band noise and part of the high-frequency products generated by the square-law detection. In the system under test it is implemented digitally, after quantization, so that it does not shape the signal presented to the quantizer. It acts instead as the anti-aliasing filter for the decimation to the $s'$ samples per symbol on which the RX-DSP operates. 

\section{Transceiver optimization}
The general optimization framework presented in Section~\ref{sec:Modeling-and-optimization} encompasses both RX-only optimization and joint TX--RX optimization, with the former used in combination with BPAM signaling, and the latter exploiting a DM  of the communication channel. The following subsections describe: the BPAM approach for TX (subsection A), RX-side optimization strategies for BPAM (subsection B),  E2E optimization techniques (subsection C), and the evaluation of algorithms complexity (subsection D).

\subsection{BPAM}\label{subsec:BPAM}

BPAM with direct detection was originally proposed in \cite{Secondini2020} based
on the observation that phase information is not completely lost after
photodetection, but can be partially recovered from the high-frequency
components of the detected signal by employing oversampling at the
receiver. In systems affected by optical noise, this approach provides
a significant sensitivity improvement over conventional unipolar PAM.

Within the general framework introduced in the previous section, BPAM
can be regarded as a particular implementation of the TX-DSP and RX-DSP
mappings. Assuming $m=2$ bits per symbol (BPAM-4), the TX-DSP first
maps the input bit vector $\mathbf{b}[k]=(b[2k],b[2k+1])^{T}$ onto
a BPAM symbol $a[k]\in\{\pm1,\pm {2}\}$. Specifically, the first bit
is mapped onto the symbol amplitude, whereas the second bit is differentially
encoded and determines its sign. The resulting BPAM-4 symbols are then
pulse-shaped to generate the transmitted sequence $x[\cdot]$, thus
implementing the mapping in (\ref{eq:f_TX}).

At the receiver, the waveform is sampled at $s'=2$ samples per symbol.
The RX-DSP computes two decision variables from the received samples.
For an ideal channel and optical matched filter, the first one is
simply obtained from the sample taken at the center of the pulse and
carries information about the transmitted amplitude. The second decision
variable is obtained by combining the sample located midway between
two adjacent pulses with the two neighboring pulse-center samples,
thereby extracting the information carried by the differential sign
while suppressing the amplitude contribution. In the notation introduced
in the previous section, the resulting decision variables can be expressed
as
\begin{align}
z[2k] & =y[2k]\label{eq:bpam1}\\
z[2k+1] & =y[2k+1]-c(y[2k]+y[2k+2])\label{eq:bpam2}
\end{align}
where the coefficient $c$ depends on the overall impulse response
(pulse shaping + matched filter). The amplitude bit $b[2k]$ is then
recovered by applying a threshold detector to $z[2k]$, whereas the
differentially encoded sign bit $b[2k+1]$ is obtained by thresholding
$z[2k+1]$.

\subsection{Receiver DSP}\label{subsec:Receiver-DSP}
\begin{table*}[h]
    \centering
    \caption{Computational complexity and memory requirements for RX-DSP implementations (with $L_{y1}=L_{y2}=L_{y}$)}
    \begin{tabular}{c|c|c}
         Method & Operations per bit & Memory  per bit\\
         \hline
         Linear eq. & $L_y$ multiplications and $L_y-1$ additions & $(2L_y-1)p$   \\
         LUT & 1 comparison & $2^{N_{\text{ADC}}L_y}$  \\
         Tree & $T_{\text{AD}}/2$ comparisons & $p T_{\text{NN}}/2$  \\ 
         ~ & $\sum_{v=1}^{V}H_{v}H_{v-1}$ multiplications& \\
          NN &$\sum_{v=1}^{V}H_{v}H_{v-1}$ additions & $p(1+\sum_{v=1}^{V}H_{v}(1+H_{v-1}))$\\
           ~ &$1+\sum_{v=1}^{V}H_{v}$ comparisons &\\ 
    \end{tabular}
    \label{tab:Cost_alg}
\end{table*}

RX-side DSP has traditionally relied on linear feed-forward and decision-feedback equalizers, whose coefficients are adapted according to minimum mean-square error criteria~\cite{Che2024}. Although highly effective for approximately linear channels, these techniques become suboptimal in short-reach IM/DD systems, where bandwidth limitations, nonlinear electro-optical components, square-law photodetection and non-Gaussian noise jointly introduce impairments that are difficult to compensate using linear processing alone~\cite{Agazzi2005,Stojanovic2017}. To address these limitations, model-based nonlinear receivers have been extensively investigated. Volterra equalizers extend linear filtering by incorporating nonlinear kernels with memory, while maximum-likelihood sequence estimation explicitly exploits channel statistics to approach optimum detection performance~\cite{Agazzi2005,Stojanovic2017,ZhangJW2022}. Although these techniques generally improve detection accuracy, their computational complexity increases rapidly with channel memory and nonlinear order, motivating numerous reduced-complexity implementations~\cite{Diamantopoulos2019,Yu2020,Taniguchi2023}. 
More recently, data-driven approaches have emerged as an attractive alternative. Neural-network receivers learn the nonlinear decision function directly from training data without requiring an explicit channel model, often achieving performance comparable to or better than model-based nonlinear equalizers while offering greater architectural flexibility~\cite{Reza2018,Xu2020cascade,DaRos2020,Bluemm2023,Schadler2021}. In parallel, memory-based detectors, including LUTs and decision-tree classifiers, have been investigated to trade computational complexity for memory usage and enable efficient hardware implementations of nonlinear detection rules~\cite{Chen2022mlse,Xie2024clut,Javadi2026,Mendonca2026}.

The following subsections describe four representative receiver detection strategies within the unified optimization framework introduced in Section~\ref{sec:Modeling-and-optimization}: linear equalizer, LUT-based  detection, decision tree, and NN. Their computational complexity and memory requirements are reported in Table~\ref{tab:Cost_alg}, and typical values are shown in Section~\ref{sec:results}.

\subsubsection{Linear equalizer}

To compensate for the distortions introduced by the transmission channel
and non-ideal analog components, the RX-DSP can be implemented as
a fractionally spaced FFE operating at $s'=2$
samples per symbol. Within the general framework introduced in Section~\ref{sec:Modeling-and-optimization},
and assuming the BPAM-4 modulation described in Section~\ref{subsec:BPAM},
this corresponds to a linear realization of the mapping in (\ref{eq:f_RX})
\begin{equation}
\mathbf{z}[k]=\mathbf{C}^{T}\mathbf{Y}[k]
\end{equation}
where $\mathbf{C}\in\mathbb{R}^{L_{y}\times2}$ is the equalizer coefficient
matrix. In this implementation, the RX-DSP parameters coincide with
the equalizer coefficients, i.e., $\boldsymbol{\theta}_{\mathrm{RX}}\equiv\mathbf{C}$.
The two components of $\mathbf{z}[k]$ provide estimates of the transmitted
squared amplitude $a^{2}[k]$ and differential sign $\varphi[k]=\sign{(a[k]a[k+1])}$,
respectively, and the transmitted bits are recovered by applying the
same threshold detectors described in the previous subsection.

The coefficient matrix $\mathbf{C}$ is selected to minimize the mean-square
error (MSE) between the equalizer output $\mathbf{z}_{k}$ and the
corresponding target decision variables $\mathbf{d}[k]=(a^{2}[k],\varphi[k])^{T}$over
a training set of $N$ input--output pairs. This can be accomplished
either iteratively, using gradient-based algorithms, or in closed
form by solving the linear least-squares (LLS) problem
\begin{equation}
\mathbf{C}_{\mathrm{opt}}=\argmin_{\mathbf{C}}\sum_{k=1}^{N}\left\Vert \mathbf{C}^{T}\mathbf{Y}[k]-\mathbf{d}[k]\right\Vert ^{2}\label{eq:LLS_problem}
\end{equation}
whose solution is
\begin{equation}
\mathbf{C}_{\mathrm{opt}}=\mathbf{R}_{\mathbf{YY}}^{-1}\mathbf{R}_{\mathbf{Yd}}\label{eq:LLS_solution}
\end{equation}
with
\begin{equation}
\mathbf{R}_{\mathbf{YY}}=\sum_{k=1}^{N}\mathbf{Y}[k]\mathbf{Y}^{T}[k],\qquad\mathbf{R}_{\mathbf{Yd}}=\sum_{k=1}^{N}\mathbf{Y}[k]\mathbf{d}^{T}[k]\label{eq:LLS_solution_correlation_matrices}
\end{equation}

The formulation above assumes a common context window for both decision
variables, spanning and integer number of symbol intervals, resulting
in a single input vector $\mathbf{Y}[k]$. While convenient for analysis
and consistent with the notation introduced in Section~\ref{subsec:End-to-end-optimization},
this assumption is not mandatory. In practical implementations, the
two decision variables may instead be generated by two independent
equalizers operating on different subsets of received samples, whose
lengths can be independently optimized to achieve the desired trade-off
between accuracy and complexity. The two equalizers therefore operate
on the input windows 
\begin{align}
\mathbf{Y}_{1}[k] & =\left(y[2k-L_{y1}^{-}],\ldots,y[2k+L_{y1}^{+}]\right)^{T}\in\mathbb{R}^{L_{y1}}\label{eq:inwin1}\\
\mathbf{Y}_{2}[k] & =\left(y[2k+1-L_{y2}^{-}],\ldots,y[2k+1+L_{y2}^{+}]\right)^{T}\in\mathbb{R}^{L_{y2}}\label{eq:inwin2}
\end{align}
which may have different lengths and are centered on samples separated
by one sampling period. The corresponding decision variables are computed
as
\begin{equation}
z[2k+i-1]=\mathbf{c}_{i}^{T}\mathbf{Y}_{i}[k],\qquad i=1,2
\end{equation}
where $\mathbf{c}_{i}$ denotes the coefficient vector of the $i$-th
equalizer. The optimal coefficients can be found by solving two independent
LLS problems, analogously to (\ref{eq:LLS_problem})--(\ref{eq:LLS_solution_correlation_matrices}).
For an ideal channel, the BPAM receiver described in Section~\ref{subsec:BPAM}
is recovered as a particular case of this formulation by choosing
$L_{y1}^{-}=L_{y1}^{+}=0$ and $L_{2}^{-}=L_{2}^{+}=1$ , corresponding
to $L_{y1}=1$ and $L_{y2}=3$ total coefficients for the two equalizers.
The computational complexity of the proposed linear receiver is dominated by the evaluation of the equalizer outputs. Since BPAM-4 conveys two information bits per transmitted symbol, the implementation based on two independent FFEs requires $(L_{y1}+L_{y2})/2$ real multiplications and $(L_{y1}+L_{y2}-2)/2$ real additions per detected bit. Memory occupation is limited to the number of coefficient with their precision. Assuming a precision of $p$ bits, the memory required by a linear equalizer is $M_{LE}=(L_{y1}+L_{y2}-1)p$.

\begin{figure}[!t]
\centering
\includegraphics[width=1\columnwidth]{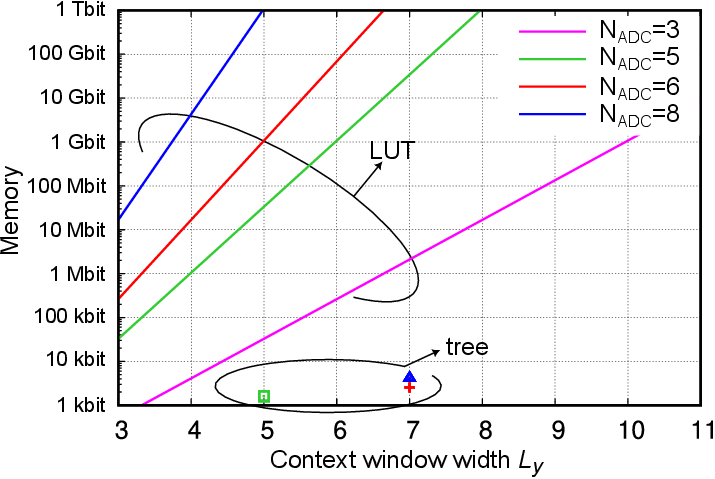}
\caption{\label{fig_memory} Memory required for LUT and decision tree storage versus context window width $L_y$, for different $N_{\text{ADC}}$ with the same color. The memory is taken from  Table \ref{tab:Cost_alg}. For classification tree, $p=8$ and the number of nodes $T_{\text{NN}}$ is that obtained in Section~\ref{subsec:sim_resu}}
\end{figure}

\subsubsection{Lookup table (LUT)}
Within the general framework introduced in Section~\ref{sec:Modeling-and-optimization},
and assuming the BPAM-4 modulation described in Section~\ref{subsec:BPAM}, we propose a LUT-based detector implementing maximum-a-posteriori probability (MAP) detection.

After the ADC, the BPAM-4 photodetected signal is described by $s'=2$ samples per symbol, and each sample by $N_{\text{ADC}}$ bits. Assuming an ideal channel and optical matched filter, according to Eqs.~(\ref{eq:bpam1})-(\ref{eq:bpam2}) three samples ($3\cdot 2^{N_{\text{ADC}}}$ bits) are sufficient to reconstruct each transmitted symbol, carrying two information bits.  However, when considering a non-ideal channel and subsystems as well as propagation along the fiber, the system introduces memory and more adjacent samples must be considered. For this reason, and similarly to the linear equalizer, the RX-DSP operates on a larger context window of  $L_y$ samples (see Eq.~(\ref{eq:cont_win})), described by $L_yN_{\text{ADC}}$ bits.  

Thus, the RX-DSP can be implemented with a negligible-computational-complexity LUT containing the transmitted bits for all possible received sequences. Given the received sequence $\mathbf{Y}[k]$ in (\ref{eq:cont_win}), and assuming an odd context window width $L_y$ with $L_y^+=(L_y-1)/2+1$ and $L_y^-=(L_y-1)/2-1$,  the LUT selects as  $k$-th and $k+1$-th transmitted bits those contained in the address specified by $\mathbf{Y}[k]$ (the address is simply its binary representation).  
The LUT should have $2^{L_yN_{\text{ADC}}}$ columns, and requires a memory of $M_{LUT}=2^{L_y N_{\text{ADC}}+1}$ bits. While the required memory explodes with $N_{\text{ADC}}$ and $L_y$, and becomes soon impracticable, the decoding has zero computational complexity, making this solution very attractive from an implementation point of view.

Figure~\ref{fig_memory} shows the memory per bit required for different $N_{\text{ADC}}$ and context windows $L_y$ with solid lines, reported in Table~\ref{tab:Cost_alg}. While increasing both parameters allows, in principle, to improve the performance---increasing $L_y$  compensates  for interferences with larger memory, while $N_{\text{ADC}}$ increases the  resolution on the received samples---, the figure shows that these numbers should be very small to allow a LUT implementation, considering 100Mbit as a benchmark for the maximum allowed memory.

In this work, we train the LUT according to the MAP strategy, which selects as decoded bits those which maximize the probability of being transmitted, given that the sequence of $L_y$ samples was received. The a posteriori probability is estimated in a training phase with $N_{\text{tr}}$ symbols, sent in the channel and received with the same setup used for transmission: for each sequence of $L_y$ received samples, the a posteriori probability of transmitted bits is estimated. The accuracy of this estimation increases with the number of appearances of each sequence.  If a sequence never occurs, the LUT is filled with the same bits decided for a \textit{similar} sequence, i.e., with minimum Euclidean distance. While, as we will show, a \textit{reasonable} number of symbols is sufficient, the accuracy of this estimation can be, in principle, increased at will, since the training is performed once and offline. 

For the sake of simplicity, the LUT-based approach has been described and implemented as a single LUT taking decision on the BPAM-4 symbol, i.e., on $2$ bits. However, the approach could be implemented with 2 smaller LUTs, one for the phase bit and one for the amplitude bit. A comparison among the two implementations, to establish the best trade-off among performance and hardware requirements, is left for future work. The same consideration applies for the decision tree in the next section.

\subsubsection{Decision tree}

The exponential memory growth associated with the LUT implementation makes impracticable to consider large $L_y$ and $N_{\text{ADC}}$, thus hampering the system performance. Therefore, we  propose to replace the LUT with a decision tree, implementing the same (or a similar) decision while drastically reducing the required memory,  at the expense of an additional computational complexity. 
The decision tree is trained using the same training dataset employed for the LUT. Tree construction relies on the Gini diversity index as the split criterion, followed by cost-complexity pruning to reduce the model complexity and improve generalization.
The decision tree is expected to provide the same performance of LUT, while being able to consider larger $L_y$ and $N_{\text{ADC}}$.
From the hardware point of view, in the inference phase, the decision tree is traversed from the root to a leaf, requiring a number of comparisons proportional to the average tree depth $T_{AD}$, while the memory required is proportional to the number of stored nodes $T_{\text{NN}}$.
Fig.~\ref{fig_memory} reports with symbols the memory requirements of some decision trees, where the number of nodes corresponds to the values  obtained in Section \ref{subsec:sim_resu}. The figure shows that, as expected, the memory required to store the decision tree is much smaller than that required for the LUT. Of course, this comes at the expense of an additional computational complexity.

\subsubsection{Neural network (NN)\label{subsec:RX-NN}}
\begin{figure}[!t]
\centering
\includegraphics[width=1\columnwidth]{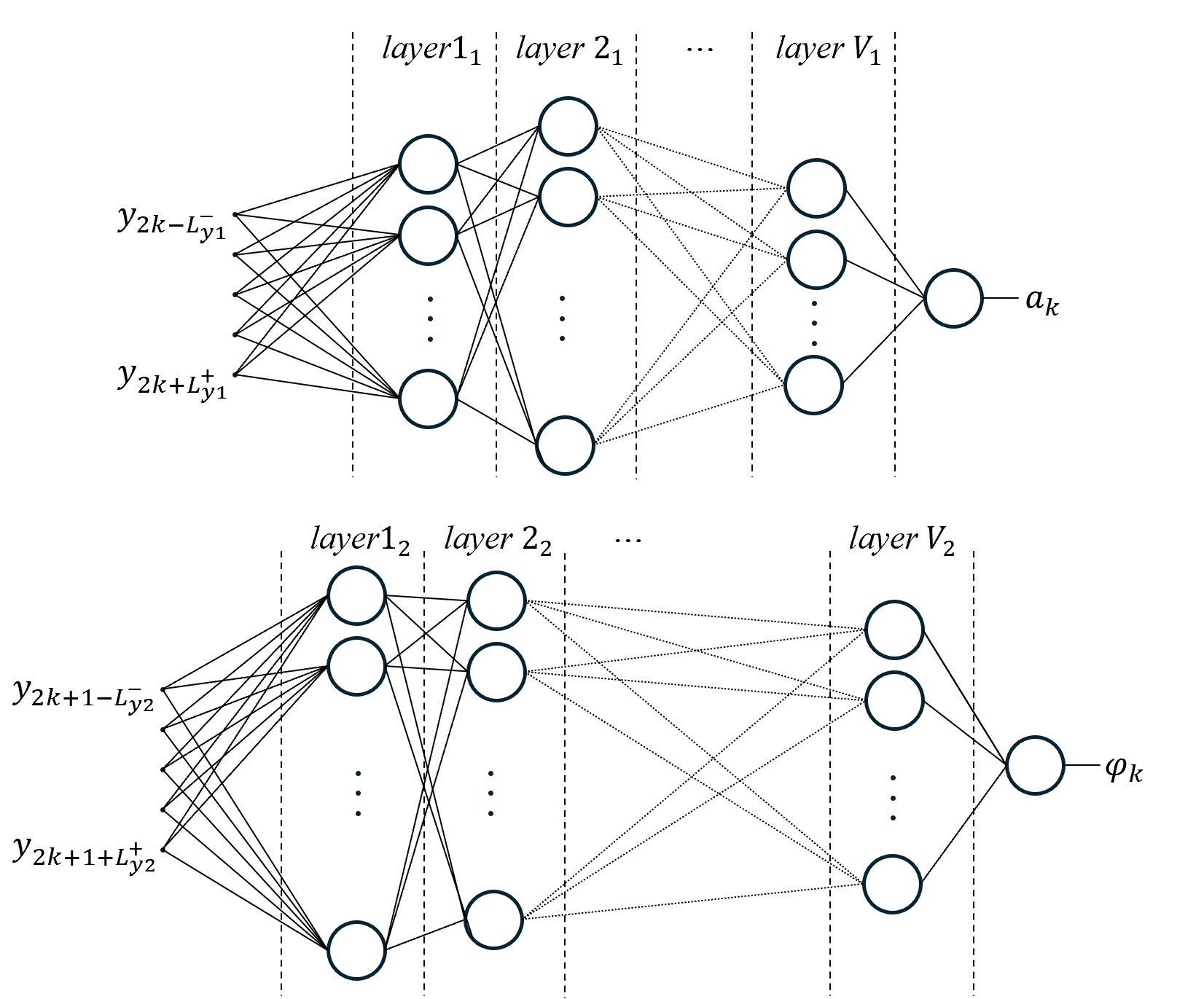}
\caption{NNs architecture for BPAM-4 amplitude and phase decisions.}
\label{fig:NN}
\end{figure}

Another approach to BPAM-4 detection is to use the ADC output samples as inputs to a NN trained to minimize the cross-entropy between the transmitted bits and their estimated posterior probabilities. With reference to the notation introduced in Section~\ref{subsec:End-to-end-optimization}, a fully connected NN provides the output variable $\mathbf{z}[k]$ as given in (\ref{eq:f_RX}) once the RX-DSP parameters $\boldsymbol{\theta}_{\mathrm{RX}}$ have been trained. 
As for the linear equalizer, the two BPAM-4 decision variables can  be generated by two independent NNs operating on different subsets of the received samples. Let the networks consist of $V_i$ hidden layers, where $i=1,2$ identifies the two received samples subset, as shown in Fig.~\ref{fig:NN}. The $v_i$-hidden layer contains $H_{v,i}$ neurons, $v_i=1,\ldots,V_i$. Each neuron is fully connected to all neurons in the preceding layer and is followed by a leaky rectified linear unit (LeakyReLU) as nonlinear activation function, except for the output layer, where a sigmoid activation is employed to produce probabilities in the interval $[0,1]$.
In particular, the two networks process the input vectors  (\ref{eq:inwin1}) and  (\ref{eq:inwin2})
where
\begin{equation}
L_{yi}=L_{yi}^{-}+L_{yi}^{+}+1,
\qquad i=1,2.
\end{equation}
The two input windows are centered on samples separated by one sampling period, consistently with the structure adopted for the two independent linear equalizers. In this case,
\begin{equation}
z[2k+i-1]=
f_{\mathrm{RX},i}
\left(
\mathbf{Y}_{i}[k];
\boldsymbol{\theta}_{\mathrm{RX},i}
\right),
\qquad i=1,2,
\end{equation}
where $z[2k]$ and $z[2k+1]$ represent the estimated a posterior probabilities of the amplitude and differential-phase bits, respectively. The RX-DSP parameters coincide with the weights and biases of the neural
networks, i.e., $\boldsymbol{\theta}_{\mathrm{RX,i}}$ contains all trainable network parameters. Therefore, the total number of parameters for the $i$-th NN is
 
\begin{equation}
\sum_{v=1}^{V_i}H_{v,i}\left(H_{v-1,i}+1
\right)+H_{V_i}+1,
\label{eq_dNN}
\end{equation}
with $H_{0,i}=L_{yi}$. Therefore, for fixed hidden-layer architectures, the parameter count and the corresponding inference memory grow linearly with the sum of the two context-window lengths, $L_{y1}+L_{y2}$. Compared with a single network jointly estimating both bits, the proposed decomposition permits the two detection tasks to use independently optimized input windows and model capacities, at the cost of duplicating part of the network structure.
The proposed NN receiver is primarily intended as a benchmark to assess the potential of RX-DSP, rather than as a directly deployable real-time architecture. Indeed, the number of operations and memory required  for its practical  implementation as a fully-connected NN---reported in Table~\ref{tab:Cost_alg}---becomes unpractical as the number of inputs, nodes, and layers increase. For sufficiently small context windows $L_y$ and ADC resolutions $N_{\text{ADC}}$, the NN mapping can be implemented as  LUT (with complexity and memory reported in Table~\ref{tab:Cost_alg}) or approximated by a compact decision tree (see Fig.~\ref{fig_memory}). For larger input spaces, hardware-oriented approximations based on pruning, quantization, or simplified NN architectures may offer a more practical trade-off between performance and implementation complexity.

\subsection{End-to-end (E2E) optimization}
 By optimizing the TX's mapping and the RX's processing functions simultaneously, E2E learning accounts for the complete system dynamics, thereby enhancing the performance of the optical link. The foundational demonstration of this model was proposed by B. Karanov et al.~\cite{Karanov2018}, who experimentally verified the approach by achieving an information rate of 42~Gb/s over a 40~km fiber link, outperforming conventional PAM with standard receiver-side equalization. 
Building upon these works, several paradigms have been developed to enhance performance. For instance, I. Roumpos et al. introduced optics-informed NNs, integrating physical optoelectronic device principles directly into the autoencoder framework. Through this approach, an experimental validation demonstrating a 48~Gb/s transmission over a 42~km fiber link was achieved~\cite{roumpos2023high}.

Recent research has focused over short-reach, high-speed links while simultaneously managing hardware complexity. In this context, D.~Li et al. reported a low-complexity receiver-side approach using a full adder-based convolutional NN equalizer for a 200~Gbaud over a 0.5~km link driven in on-off keying by a high-bandwidth integrated ring modulator~\cite{li2026low}. To bypass the need for an accurate digital surrogate during training, Z. Li et al.~\cite{li2024model} trained the autoencoder directly on the physical link, enabling a channel-model-free E2E architecture that embeds the real features of the transmitting medium. In their work, the transceiver dynamically refines constellation distributions to optimize communication performance over physical channels, experimentally achieving a data rate up to 349.2~Gb/s over a 0.5~km single-mode fiber link.

In this section, we present two examples of E2E optimization: one based on fully connected NNs employing a physics-inspired surrogate model, and the other based on a transformer architecture.

\subsubsection{End-to-end  neural network (E2E NN)}
A possible implementation of E2E optimization uses fully-connected  with few layers NNs both at the TX and the RX. 

For the TX, the input is the bit context window $\mathbf{B}[k]$ of \eqref{eq:f_TX} and the output is the vector $\mathbf{x}[k]$. Hidden layers use a LeakyReLU activation and the output layer a $\tanh$, which bounds the drive to the allowed $[-V_\pi,+V_\pi]$ swing; the fixed TX digital filters then complete the pulse shaping. The short window keeps the predistorter representable, eventually, by the LUT of the DSP-simplification step. No signaling structure is imposed---no level spacing, no differential precoding, no fixed pulse shape---so the NN is free to tailor the modulation format itself. 

The RX-NN is the same described in Section~\ref{subsec:RX-NN}.
 

The TX-DSP and RX-DSP are jointly trained in an autoencoder-like framework using the Adam optimizer \cite{Kingma2017}. 
The optical noise level adopted during training plays a critical role. Training at a very low OSNR may significantly slow down or even prevent convergence, whereas training at a very high OSNR may lead to solutions that closely fit the noiseless channel response but generalize poorly at the target operating OSNR, similarly to a zero-forcing equalizer. Since the receiver operates at $s'=2$ samples per symbol, the TX has one additional temporal degree of freedom per symbol. Consequently, the optimization autonomously identifies the most suitable signaling format—ranging from conventional symbol-rate bipolar pulses \cite{Secondini2020} to double-symbol-rate half-symbol signaling, or intermediate solutions—according to the bandwidth constraints of the transmission link.

The complexity and memory of each NN is reported in Table~\ref{tab:Cost_alg}. For the RX, the same considerations for RX-NN in Section~\ref{subsec:RX-NN} hold. Conversely, since the TX context window is smaller, as we will show, the TX-NN can be replaced with a LUT (in this case, the memory per bit required for the LUT is $2^{L_b}(sN_{\text{DAC}})/m$).


\subsubsection{End-to-end transformers (E2E Transf.)} 


To capture long-range inter-symbol dependencies across the complete communication chain, the TX and receiver DSPs can be implemented using transformer-based architectures~\cite{vaswani2017attention}. Unlike the fully connected E2E implementation described above, the transformer processes an entire sequence of symbols jointly and dynamically weighs the contribution of different temporal positions through the multi-head self-attention mechanism~\cite{ahmed2023transformers}.
The TX-Transformer maps an input bit sequence $b[\cdot]$ into a continuous, oversampled (at a rate of $s$ samples per symbol) and neural-pulse shaped electrical waveform $x(t)$, while pre-distorting the signal against subsequent component non-linearities and bandwidth limitations. Prior to the standard multi-head attention mechanism of the Transformer, the Gray-code mapping symbol sequence $\mathbf{X}[k]$ is projected through a Volterra layer where $k \in \{0, 1, ..., L_{\mathrm{seq}}-1\}$ denotes the discrete symbol time index, $L_{\mathrm{seq}}$ is the length of the sequence, $B_s$ is the batch size, and $d_\text{model}$ is the width of the neural layers \cite{Roheda2024Volterra}. This Volterra layer provides the Volterra tensor  $\mathbf{H}_{\mathrm{Volt}} \in \mathbb{R}^{B_s \times L_{\mathrm{seq} }\times d_{\mathrm{model}}}$ as a combination of linear kernels and second-order non-linear memory structures that embed a physical prior into the network to reduce the required depth of neural layers and the convergence time for training.

The Volterra tensor $\mathbf{H}_{\mathrm{Volt}}$ enters a $3$-Layer $8$-head self-attention transformer stack whose output is ~\cite{ahmed2023transformers}
\begin{equation}
\mathbf{X}_{\mathrm{transf}} = f^{(3)}_{TX}(f^{(2)}_{TX}(f^{(1)}_{TX}(\mathbf{H}_{\mathrm{Volt}}))) \in \mathbb{R}^{B_s \times L_{\mathrm{seq}} \times d_{\mathrm{model}}},
\label{eq-li-43}
\end{equation}
\noindent where $f^{l}_{TX}(\cdot)$ represents a feed-forward neural network with a width of $4\times d_{\mathrm{model}}$ to execute dense feature extractions, which characterizes the linear/nonlinear distortions of optoelectronic components as well as their complex interplay with bandwidth limitations and AWGN, and thus accordingly, pre-distorting the signals at TX-DSP.
Instead of applying standard Nyquist-pulse shaped filters, here we used a convolutional layer with a kernel size and stride equivalent to an upsampling rate $s$. This layer adaptively learns the optimal interpolation basis functions while expanding the discrete temporal sequence to the $s$-samples-per-symbol sampling domain. The upsampled tensor is subsequently routed
through two hidden convolutional layers engineered
with extended kernel spans. These heavy temporal kernels deliberately overlap adjacent samples to neutralize localized residual ISI and smooth out high-frequency discontinuities. Interleaved with Gaussian error linear unit ($\text{GELU}$) activations and batch normalization for stabilized feature propagation, ultimately generating a smoothly-varying, distortion-resilient neural-pulse shaped output $\mathbf{X}_{\text{neural}} \in \mathbb{R}^{B_s \times L_{\mathrm{seq}} \times s}$.
Finally, the DAC model is applied as described in Section~\ref{subsec:dac}.

At the RX, the ADC output sequence is projected onto the same feature space and processed by a second transformer stack. The self-attention layers exploit the temporal correlation among the received samples and provide a nonlinear equalization and detection function over a context window that may span the entire processed sequence. The final layer produces the posterior probabilities of the transmitted bits,
\begin{equation}
f_{\mathrm{RX}}^{\mathrm{Transf}}
\left(
\mathbf{Y};
\boldsymbol{\theta}_{\mathrm{RX}}
\right),
\end{equation}

\noindent where $\mathbf{Y}$ denotes the received sample sequence and $\boldsymbol{\theta}_{\mathrm{RX}}$ collects the trainable receiver parameters. Hard bit decisions are obtained by thresholding the corresponding probabilities.

The TX and RX transformers are jointly optimized by minimizing the binary cross-entropy between the transmitted bits and the receiver soft outputs. During training, the channel parameters remain fixed, while the gradients are propagated through the complete differentiable communication chain. 

\subsection{Algorithm complexity in DSP hardware\label{subsec:alg_compl_ASIC}}

In modern optical communication systems, the DSP is implemented as an application-specific integrated circuit (ASIC). To evaluate the complexity of digital circuits independently from manufacturing technology, the gate equivalent (GE) is used as a standard metric. 1 GE is defined as the physical area of a 2-input NAND gate (4 transistors in standard CMOS technology) and is related to silicon area,  power consumption, and critical path delay, all key elements for hardware cost.

In this paper, we estimate the DSP computational complexity estimating a normalizing GE from the number of multiplications $n_{\text{mult}}$, additions $n_{\text{add}}$, and comparisons $n_{\text{comp}}$ performed for information bit. All is done with fixed-point arithmetic, as the circuitry for floating-point would not allow to satisfy the speed requirements~\cite{liu2008embedded}. A standard adder, as a carry-lookahead adder, scales its area quasi-linearly with the number of bits involved $p$. A comparator is structurally similar to a subtractor: it uses 2's complement math to subtract the numbers and simply checks the sign bit and zero flag, throwing away the rest of the summation bits. A multiplier is essentially an array of dozens of adders working together to accumulate partial products having an area scaling quadratically with $p$.  

Overall, we estimate the computational complexity per information bit of the proposed DSP techniques from the number of relative GE counts of typical $p=8$-bit implementations, as~\cite{parhamicomputer}

\begin{equation}
     12n_{\text{mult}} + 1.2n_{\text{add}} + 1n_{\text{comp}}.
     \label{eq:complexity}
\end{equation}

\section{Results and discussion\label{sec:results}}
\subsection{System setup}\label{subsec:setup}

The system setup is described in Fig.~\ref{fig:system_overview} and in Section~\ref{sec:System-overview}. Here, we provide the specifications of the system under test, either in simulation or in laboratory experiments.

The TX-DSP generates a $R_b=40$~Gb/s digital signal with $80$~Gsamples/s, either using BPAM-4 for RX-side optimization only or using the signal generated by the E2E optimization. The BPAM-4 signal is a $20$~GBd signal with $s=4$ samples/symbol modulating a time-domain raised-cosine pulse shape with roll-off $0.85$  and $m=2$~bits/symbol. A digital linear (in dB vs frequency) pre-emphasis of $6$~dB at $20$~GHz is applied to compensate for lab devices bandwidth limitations. At the end of the TX-DSP chain, a digital $20$~GHz rectangular filter is used to limit the signal bandwidth. The DAC is a  commercial device with sampling rate $80$~Gsamples/s, with a resolution of $6$ bits, and $35$~GHz analog 3~dB bandwidth, emulated in our simulations as an ideal 6-bits quantizer. The electrical driver, with 45~GHz bandwidth and maximum output voltage $3$~V, drives a MZM with 35~GHz bandwidth, $V_\pi=5$~V and $10\text{log}_{10}{\varepsilon_r}=$~30dB.  The MZM modulates an external cavity laser source at 1550~nm, and the optical signal is sent into the fiber. In simulations, the MZM is emulated as a push-pull device  with $30$~dB extinction ratio  and driven by a peak voltage equal to $0.6 V_\pi$, see Section~\ref{subsec:mzm}. The fiber is a $10.238$~km  single-mode fiber (dispersion $\beta_2=-21.7$~ps$^2$/km, $\alpha=0.2$~dB/km, $\gamma=1.3$~W$^{-1}$km$^{-1}$), simulated with the split-step Fourier method. After the fiber, the desired OSNR is obtained by a standard ASE noise loading technique using an erbium-doped-fiber-amplifier. Next, the signal passes through an LCoS-based  OBPF as described in Section~\ref{subsec:optfilt} with $B=0.22$~nm, and $B_{\mathrm{otf}}=18$~GHz, and is received with a $50$~GHz photodetector, both also included in the simulations as in Section~\ref{sec:analog-modeling}.  The electrical signal is received with an 8-bits ADC with $33$~GHz bandwidth, and $200$~Gsamples/s. The ADC is emulated in the simulations as an ideal quantizer. Additionally, to study the system behavior with a lower-resolution ADC, we further digitally quantize the signal to $N_{\text{ADC}}$ resolution bits.  Next, a digital filter with Gaussian shape and bandwidth $10$~GHz is applied taking into account hardware bandwidth limitations, followed by RX-DSP with $s'=2$ samples per symbols. Finally, the BER is estimated comparing transmitted and received bits. 

Simulations and experimental results are given as a function of the OSNR, related to the SNR per bit ${E_b/N_0}$ as   $\mathrm{OSNR}_{\mathrm{dB}}={(E_b/N_0)}_{\mathrm{dB}}+10\log_{10}{(R_b/(m \,12.5\text{GHz}))}$. 

\subsection{Simulation results}\label{subsec:sim_resu}

\begin{figure}[!t]
\centering
\includegraphics[width=1\columnwidth]{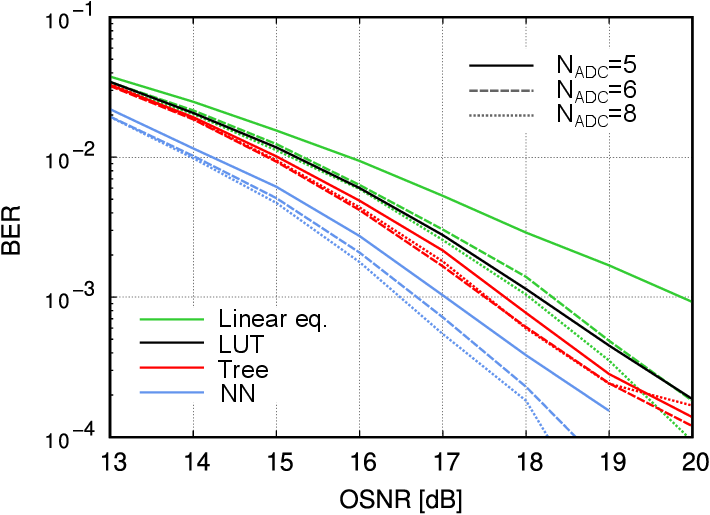}
\caption{\label{fig:fig_results_RXDSP_sim} BER versus OSNR for different RX-DSP detection strategies with BPAM-4, and for different $N_{\text{ADC}}$, obtained with simulations.}
\end{figure}

\begin{figure}[!t]
\centering
\includegraphics[width=1\columnwidth]{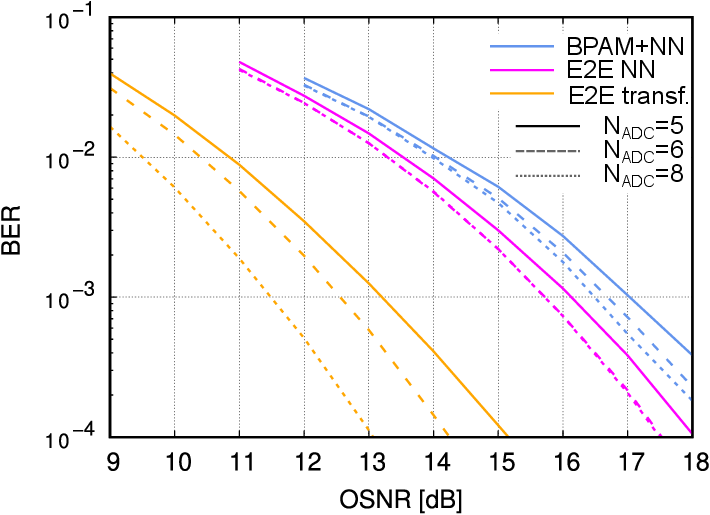}
\caption{\label{fig_results_e2e} BER versus OSNR with E2E techniques, compared with BPAM-4 with RX-NN, obtained with simulations.}
\end{figure}

\begin{figure}[!t]
\centering
\includegraphics[width=1\columnwidth]{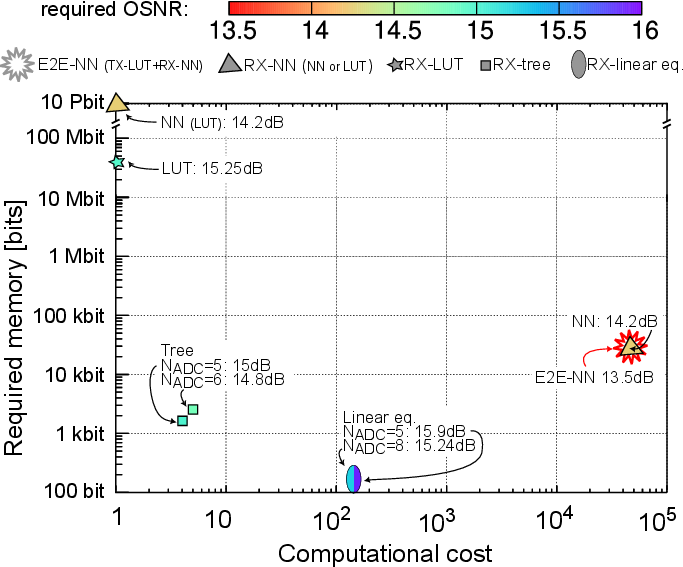}
\caption{\label{fig_hardware_perfo} Memory versus computational cost for BPAM-4 with different RX-DSP strategies and E2E-NN, reporting the OSNR required for $\text{BER}=10^{-2}$.}
\end{figure}

This section analyses the performance and complexity of the proposed TX- and RX-side strategies, obtained through simulations. In all cases presented here, the training is performed at the same OSNR used for the test phase, with $N_{\text{tr}}=2^{20}$ symbols, while the test is done with $N_{\text{tr}}=2^{18}$ symbols. 

Figure~\ref{fig:fig_results_RXDSP_sim} compares BER versus OSNR for BPAM-4 with different RX-DSP strategies (with different colors), and considering a different number of ADC bit $N_{\text{ADC} }$ (with different line styles). The linear equalizer performance is obtained with the context window $L_y=11$. The LUT is implemented with  $N_{\text{ADC} }=5$ and $L_y=5$, entailing a memory of $\approx 30$~Mbit, as larger values are considered impracticable. The decision tree is implemented with $L_y=5$ or $L_y=7$ for  $N_{\text{ADC} }=5$, and $N_{\text{ADC} }=6,8$, respectively. The choices for $N_{\text{ADC} }=5,6$ ensure a good trade-off between performance and complexity, while $N_{\text{ADC} }=8$ is for comparison with other techniques. The number of nodes, average depth, and maximum depth are: 405, 8, 15  for $N_{\text{ADC} }=5$; 637, 9, 19 for $N_{\text{ADC} }=6$; and 679, 9, 18 $N_{\text{ADC} }=8$. Finally, the NN is implemented with $L_y=11$, $V_1=V_2=3$ number of hidden layers for amplitude and phase decision, and $H_{1,i}=32$, $H_{2,i}=64$, and $H_{3,i}=16$ neurons  with $i=1,2$, to assess the potential performance of RX-DSP. Such a structure has 3553 trainable parameters per NN. 
 First, Fig.~\ref{fig:fig_results_RXDSP_sim} shows that low ADC resolution, e.g., $N_{\text{ADC} }=5$, affects the linear equalizer more that the other approaches; while $N_{\text{ADC} }=6$ provides almost the same performance of $N_{\text{ADC} }=8$. Interestingly, the decision tree with $N_{\text{ADC} }=8$ has slightly worse performance than $N_{\text{ADC} }=6$; this is due to a reduced accuracy in the training of the $N_{\text{ADC} }=8$-tree, which indeed has a very similar number of nodes and depth. For a similar reason, the tree with the same parameters of the LUT, $N_{\text{ADC} }=5$ and $L_y=5$, though expected to have the same performance, achieves lower BER. This suggests that a better training of the LUT may provide additional gain. Next,  Fig.~\ref{fig:fig_results_RXDSP_sim} shows that both LUT and decision tree perform better than the linear equalizer, with up to $1$~dB  gain at $\text{BER}=10^{-2}$. Finally, the NN,  serving as a benchmark for the potential of RX-side DSP optimization, shows that an additional gain of $\approx1$~dB might be obtained.

Figure~\ref{fig_results_e2e} reports the BER versus OSNR of the proposed E2E techniques, comparing with BPAM-4 with RX-NN. 
The E2E-NN is implemented with the following parameters. The TX-NN has a context window of $L_b=10$ bits, a single hidden layer with 16 neurons,  $s=2$ output samples, and uses $210$ parameters. Thus, the TX-NN can be implemented as a LUT with $2^{10}$ entries and memory $2^{11}$, removing all multiplications at runtime.  The RX-NN is implemented as in  Section~\ref{subsec:RX-NN} with the same parameters as NN. 
The E2E transformer is implemented using three encoder layers, eight attention heads and following parameters: $L_{\mathrm{seq}}=64$, $d_{\mathrm{model}}=128$, and $B_V=128$. Moreover, the optical noise level is randomly varied over the target operating range at each training iteration, so that the learned transceiver is robust to different OSNR.

Fig.~\ref{fig_results_e2e} shows that E2E-NN outperforms RX-NN by $\approx 0.8$~dB at $\text{BER}=10^{-2}$, while E2E optimization based on transformers provides $\approx 3$~dB of additional gain. The $\approx 0.8$~dB improvement provided by E2E-NN, though \textit{smaller} than the one provided by transformer, is particularly attractive since it is obtained replacing the BPAM-4 scheme with a negligible-complexity, small memory, LUT. Conversely, the huge improvement of E2E transformers highlights the great potential of E2E implementation, with almost $6$~dB of gain with respect to BPAM-4 when received using a linear equalizer ($N_{\text{ADC}}=8$). Additionally, the figure shows that, differently from the other approaches, transformers perform significantly better when using  $N_{\text{ADC}}=8$ rather than $N_{\text{ADC}}=6$.

Figure~\ref{fig_hardware_perfo} compares the proposed techniques for  hardware requirements and performance, except for the  E2E-transformer-based optimization whose complexity is too high. Hardware requirements are measured as  required storage memory and computational complexity, measured according to Eq.~(\ref{eq:complexity}) and Table~\ref{tab:Cost_alg}.  The parameters for the various implementations are those used for Fig.~\ref{fig:fig_results_RXDSP_sim}. The performance is given as OSNR required to achieve $\text{BER}=10^{-2}$, when implemented with BPAM-4. For the RX-DSP NN ($N_{\text{ADC}}=5$), we reported  the requirements for both its direct implementation as a NN and its implementation as a LUT (labeled as "NN (LUT)" in the figure): both implementations are not feasible and additional research is required to implement the NN with reasonable complexity, for example using decision trees. For the E2E-NN, we considered an implementation with LUT at the TX and NN at the RX, thus   increasing the required memory by a negligible value. Hence, the E2E-NN and the RX-NN are superimposed for hardware requirements, with the former performing better. The other  techniques offer different trade-off between complexity and performance, with the decision tree offering good performance with low memory requirements and computational complexity.

\subsection{Experimental results}
The experimental results were obtained using waveforms generated through numerical simulations and uploaded to the DAC. The DAC memory was limited to $2^{19}$ samples, with a vertical resolution of $N_{\text{DAC}}=6$ bits. At a sampling rate of 80~GSa/s, the DAC generated a periodic continuous-time waveform with a duration of approximately $6.5~\mu\mathrm{s}$.

The DAC output amplitude was set to 300~mV, resulting in a maximum voltage of 3~V at the input of the MZM after electrical amplification. Bipolar signals with uniformly distributed symbols and symmetric amplitude levels have zero mean. Therefore, the MZM bias point could be controlled by minimizing its optical output power. This simple bias-control procedure proved essential during the measurements, as the performance of BPAM-4 was found to be highly sensitive to deviations from the optimum bias point.

At the receiver, the signal was acquired using a real-time oscilloscope with a 33~GHz analog bandwidth and a sampling rate of 200~GSa/s. Each acquisition consisted of $2\times10^6$ samples with an ADC resolution of $N_{\text{ADC}}=8$ bits. The acquired traces were subsequently transferred to a computer and processed offline.
The waveform was then resampled and temporally aligned with the transmitted data sequence to emulate for clock recovery.
Hence, the sampling rate was reduced to $s'=2$ samples per symbol and filtered using a 10~GHz 3dB bandwidth Gaussian filter.
The processed samples were used as the common input to all four RX-DSP techniques: linear equalizer, LUT, tree, and NN.

\begin{figure}[!t]
\centering
\includegraphics[width=1\columnwidth]{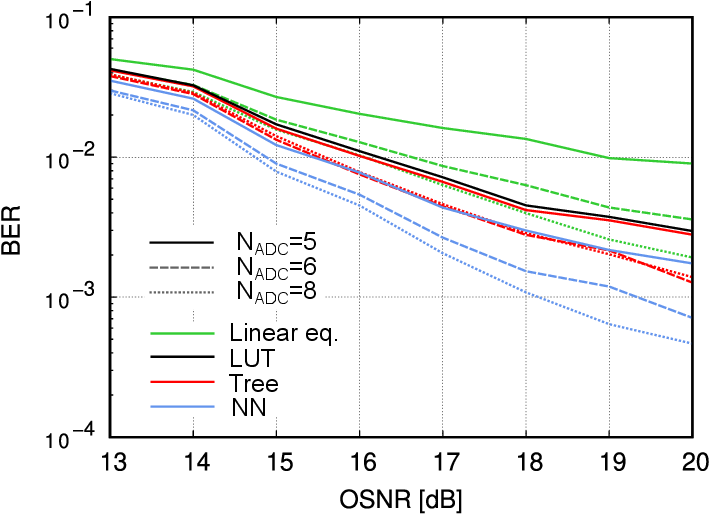}
\caption{\label{fig_results_lab} BER versus OSNR for different detection strategies with BPAM-4, obtained in laboratory experiments.}
\end{figure}

Figure~\ref{fig_results_lab} shows  experimental results for the BER as a function of the received OSNR. The training of RX-DSP techniques (LUT, tree, NN) is performed with $2^{21}$ symbols with $\text{OSNR}=16$~dB. 

The comparison between experimental results in Fig.~\ref{fig_results_lab} and simulations result in Fig.~\ref{fig:fig_results_RXDSP_sim} shows an excellent laboratory implementation, with a penalty of approximately 0.8--1~dB at $\text{BER}=10^{-2}$.
The experimental measures are in good agreement with simulations thus validating all the techniques, with the only exception being the linear equalizer with $N_{\text{ADC}}=5$ which suffers an higher implementation penalty. E2E validation is left for future work.

\section{Conclusion}

This paper proposed a comprehensive optimization framework for short-reach direct-detection optical communication systems, a key technology for next-generation data-center interconnects. After introducing the proposed framework and the underlying system and subsystems models, we investigated RX-side DSP optimization for BPAM-4 signaling, which carries information on both the amplitude and phase of the optical field. We describe and compare practical detection techniques based on linear equalizers, LUTs, decision trees, and deep learning models, explicitly considering their hardware implementation. Next, we explored E2E optimization techniques based on feedforward NNs or transformers.

The proposed techniques were tested in simulation and in a laboratory experiment, transmitting $40$~Gbit/s in a system with strong bandwidth limitations, low-to-high ADC and DAC resolutions, MZM non-idealities,  $10$~km fiber, and variable OSNR. Key simulation results showed that (i) RX-DSP based on LUT or decision tree have excellent hardware requirements and perform better or similar to the low-complexity linear equalizer, (ii) RX-DSP based on NN further improves the performance by $1$~dB but its current implementation remains challenging, (iii) E2E-NN allows to improve the performance by $0.8$~dB with respect to RX-NN, adding a small TX LUT, (iv) transformer-based E2E optimization provides a huge performance improvement of approximately $3$~dB and $6$~dB with respect to E2E-NN or linear equalizer, respectively. Although the latter has a prohibitive complexity, its performance highlights the huge capabilities of E2E optimization for these systems. Laboratory results confirm the results obtained with simulations for RX-DSP, with 1~dB implementation penalty.

Overall, the proposed framework provides both practical low-complexity receiver solutions and benchmarks for future machine-learning-based and E2E optimization techniques for short-reach direct-detection optical communication systems.
\section*{Acknowledgment}
The authors would like to thank Alessandro Cioni for his support in lab measures.

\ifCLASSOPTIONcaptionsoff
  \newpage
\fi

%

\bibliographystyle{ieeetr}
\bibliography{bibliography}

\begin{IEEEbiographynophoto}{Luca Potì}
 is full professor at the Universitas Mercatorum, Rome, Italy, and Head of the Research Area “High Capacity and Secure Optical Communications” with the Interuniversity National Consortium for Telecommunications (CNIT) at the Photonic Networks and Technologies Nat’l Lab. He is author of more than 400 papers and 50 patents. He coordinated and took part into several national, European, and international research projects.
His research interests are in the area of high capacity optical transmission systems, photonic integrated devices, quantum communications and physical layer security. 
\end{IEEEbiographynophoto}

\end{document}